# Direct PET Image Reconstruction Incorporating Deep Image Prior and a Forward Projection Model

Fumio Hashimoto and Kibo Ote

*Abstract*— **Convolutional neural networks (CNNs) have recently achieved remarkable performance in positron emission tomography (PET) image reconstruction. In particular, CNN-based direct PET image reconstruction, which directly generates the reconstructed image from the sinogram, has potential applicability to PET image enhancements because it does not require image reconstruction algorithms, which often produce some artifacts. However, these deep learning-based, direct PET image reconstruction algorithms have the disadvantage that they require a large number of high-quality training datasets. In this study, we propose an unsupervised direct PET image reconstruction method that incorporates a deep image prior framework. Our proposed method incorporates a forward projection model with a loss function to achieve unsupervised direct PET image reconstruction from sinograms. To compare our proposed direct reconstruction method with the filtered back projection (FBP) and maximum likelihood expectation maximization (ML-EM) algorithms, we evaluated using Monte Carlo simulation data of brain [$^{18}$F]FDG PET scans. The results demonstrate that our proposed direct reconstruction quantitatively and qualitatively outperforms the FBP and ML-EM algorithms with respect to peak signal-to-noise ratio and structural similarity index.**

## I. Introduction

POSITRON emission tomography (PET) is one of the major functional imaging modalities that can screen for cancer and Alzheimer's disease; however, due to the inherent involved physical degradation processes, the quality of PET images is often poor when compared with other tomography modalities such as X-ray computed tomography and magnetic resonance imaging [1]. Low-dose PET imaging has recently been proposed to reduce patients' radiation exposure, which further increases issues of statistical noise. Therefore, various different approaches have been studied, including post-denoising and image reconstruction, to improve the quality of low-dose PET images for the early detection of small lesions, as well as for quantitative analysis.

In post-denoising methods, image-guided filter [2], non-local means filter [3,4], block matching filter [5], and wavelet denoising [6] have been proposed to achieve better PET image quality than when using Gaussian filtering. In image reconstruction algorithms, various iterative image reconstructions have been proposed that use various prior models, such as anatomical information [7-9] and edge preservation [10-12]. However, these iterative reconstructions require a handcrafted prior distribution of the PET images using the patients' anatomical information, or artificial prior information, such as locally monotonic regions.

In recent years, deep learning has been widely used in medical imaging fields [13-15] and has achieved improved performance for PET image denoising [16-18], and image reconstruction [19-21]. In particular, convolutional neural network (CNN)-based direct PET image reconstruction, which directly generates the reconstructed image from the sinogram, may be applicable to PET image enhancement because it does not require image reconstruction algorithms, which can produce some artifacts [22-25]. However, these deep learning-based direct PET image reconstruction algorithms require a large number of high-quality training datasets. In clinical practice, it is necessary to consider the effect on PET image quality when there are domain gaps between the training and testing datasets resulting from other radioactive tracers and PET scanners. To overcome the aforementioned challenges, Ulyanov et al. proposed the deep image prior (DIP), which uses the CNN structure itself as prior information [26]. The DIP has been successfully applied not only to natural images but also PET images. Anatomical information, as well as static PET images, has been used as the CNN input to improve the PET image quality over the original DIP framework [27-30].

In this study, we propose a direct PET image reconstruction method that incorporates the DIP framework. Our proposed method incorporates a forward projection model into a loss function to achieve unsupervised direct PET image reconstruction from sinograms. We quantitatively compared the results of the proposed method with those of conventional iterative reconstruction algorithms using Monte Carlo simulation data from brain [$^{18}$F]FDG PET scans.

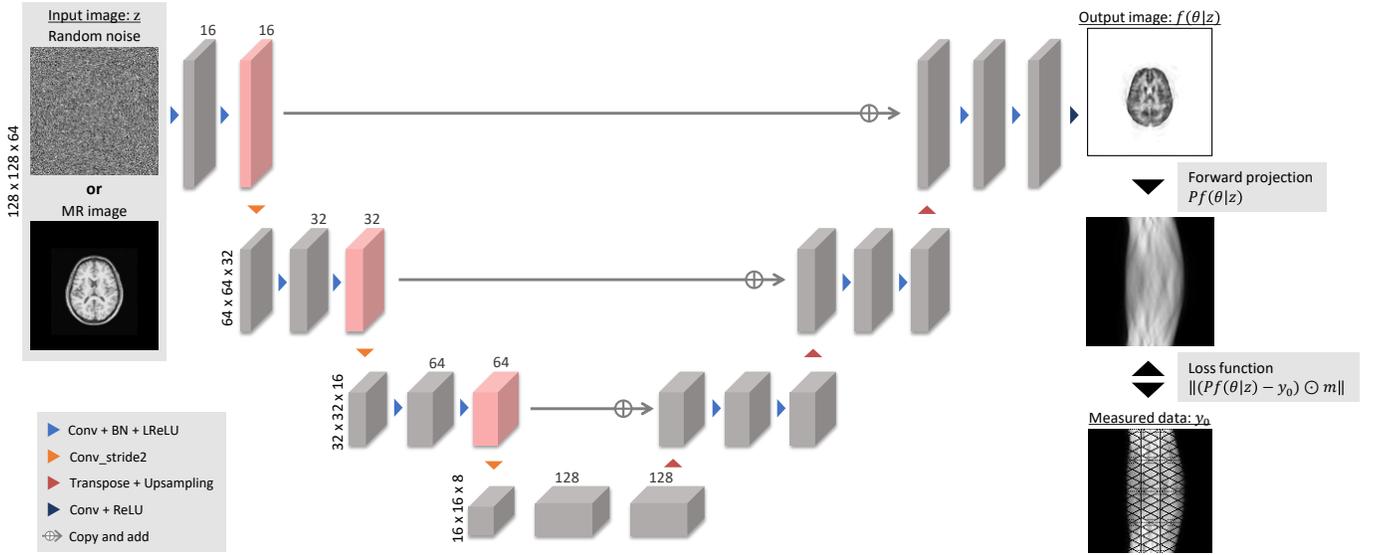

Fig. 1. Overview of the proposed direct PET image reconstruction. First, a reconstructed image was output from the CNN. Then, the L2 loss was calculated with the measured data and the estimated projection data were calculated from the CNN output.

## II. METHODOLOGY

### A. PET Forward Projection Model

In general, the PET forward projection model can be expressed such that the projection data $y \in \mathbb{R}^{M \times 1}$ are related to the spatial distribution of the radioactive tracer $x \in \mathbb{R}^{N \times 1}$ through an affine transformation, as follows:

$$y = Px, \quad (1)$$

where $P \in \mathbb{R}^{M \times N}$ is a projection matrix that denotes the contribution of each voxel to each line of response (LOR). $M$ and $N$ are the number of LOR and voxels in the image space, respectively. The projection matrix can be calculated using a rotation-based method [31].

### B. Proposed Direct PET Image Reconstruction Method

An overview of the proposed direct PET image reconstruction is shown in Fig. 1. In the proposed method, the reconstructed image $x$ is calculated using the DIP framework as follows:

$$x = f(\theta|z), \quad (2)$$

where $f$ represents the CNN, $\theta$ is the CNN weights that are the parameterization of the reconstructed image $x$, and $z$ is a prior vector of the CNN input. In the DIP framework, the implicit prior generated by indirect parametrization of the CNN weights can be used for regularization.

In this study, to calculate the reconstructed image directly, we introduce the forward projection model $P$ in the loss function:

$$\theta^* = \underset{\theta}{\mathrm{argmin}} \|(Pf(\theta|z) - y_0) \odot m\|, \quad (3)$$

$$x^* = f(\theta^*|z), \quad (4)$$

where $y_0 \in \mathbb{R}^{M \times 1}$ represents the measured projection data, $\odot$ is the Hadamard product, and $m \in \{0,1\}^{M \times 1}$ represents gaps between the detector modules, in correspondence with a binary mask. In general, the detector gaps are interpolated in the sinogram space; however, we applied the inpainting task from the DIP framework to calculate the loss function using only the measured data in this study. In this study, we used random noise and magnetic resonance (MR) images as the prior input vector $z$. The limited-memory Broyden–Fletcher–Goldfarb–Shanno (L-BFGS) algorithm was used for network training [32]. It is a quasi-Newtonian method that functions by considering the approximate Hessian matrix.

### C. Network Architecture

In this study, we used a 3D U-Net-based structure. The architecture is summarized in Fig. 1, and contained encoding (left part) and decoding (right part) segments. The encoding segment is a typical CNN structure, with repeated application of two $3 \times 3 \times 3$ convolution layers, each followed by batch normalization (BN), activation of a leaky rectified linear unit (LReLU), and one $3 \times 3 \times 3$ convolution layer with a stride of two for downsampling. The number of feature maps doubled at each downsampling layer. The decoding segment comprised one $3 \times 3 \times 3$ transpose convolution layer and trilinear upsampling layers added with the corresponding feature map from the encoding segment. It also contained two $3 \times 3 \times 3$ convolution layers, each followed by the BN and LReLU activation. Finally, we activated the $1 \times 1 \times 1$ output convolution layer using a ReLU activation.

The learning rate was set to 0.1. The proposed method was calculated using a graphics processing unit (NVIDIA Quadro RTX 8000 with 48 GB of memory) and Pytorch 1.7.1 (https://pytorch.org/).

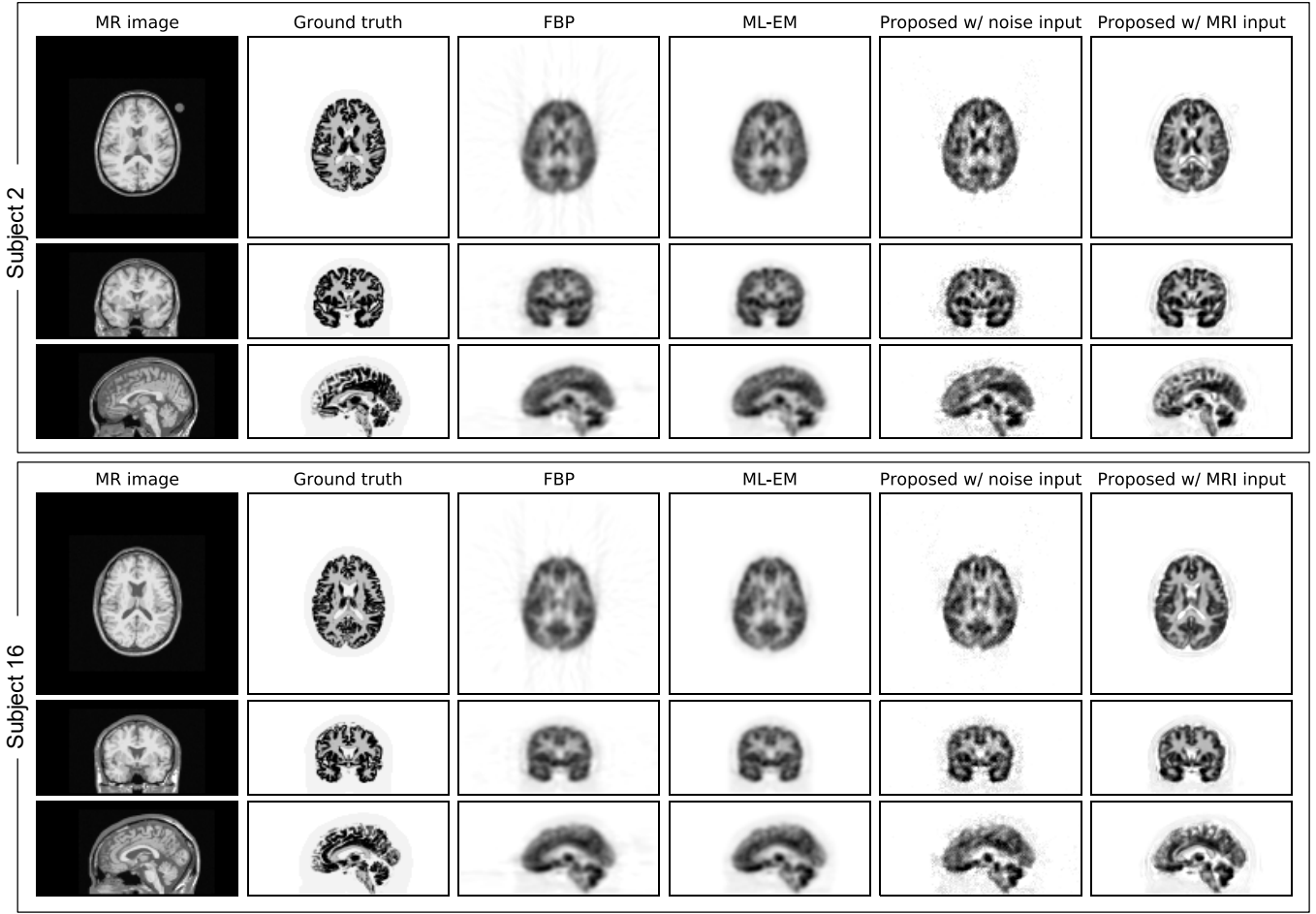

Fig. 2. Three orthogonal slices of the reconstructed images using different reconstruction algorithms. From left to right, the columns correspond to the MR images, the ground truth, reconstructed images obtained by FBP and ML-EM with Gaussian filtering of $\sigma = 1$ voxel, and the proposed direct reconstruction with random noise and MRI input.

## III. EXPERIMENTAL SETUP

### A. Simulation dataset

We employed 20 brain phantoms from BrainWeb (https://brainweb.bic.mni.mcgill.ca/brainweb/) and created the projection data using a Monte Carlo simulation. The contrast of radioactivity between the gray matter, white matter, and cerebrospinal fluid was set to 1 : 0.25 : 0.05 based on the [$^{18}$F]FDG kinetics. The attenuation coefficients of the soft tissue and bone were set to 0.00958 mm$^{-1}$ and 0.0151 mm$^{-1}$, respectively.

The following arrangement of the brain-dedicated PET scanner was used in this study. A detector ring with a diameter of 486.83 mm was constructed with 28 detector units in the ring direction and 4 in the axial direction. Each detector unit had a 16 × 16 array of cerium-doped lutetium–yttrium oxyorthosilicate (LYSO) crystals. The size of each LYSO crystal was 3.14 mm × 3.14 mm × 20 mm. The image size was 70 × 128 × 128 voxels with a voxel size of 3.221 × 3 × 3 mm$^3$. The list data from the 3D PET acquisition was transformed into 2D PET sinogram format using a single-slice rebinning method, with a maximum ring difference of ±15. The sinogram had a total of 35.88 ± 1.59 M counts for each subject. For simplicity, scatter events were removed from the list data. Attenuation correction was applied to the sinogram before image reconstruction.

### B. Evaluation

We compared the proposed direct reconstruction method with filtered back projection (FBP) using a Hanning filter and Gaussian filtering of $\sigma = 1$ voxel, and maximum likelihood expectation maximization (ML-EM) with 100 iterations and Gaussian filtering of $\sigma = 1$ voxel. To evaluate the reconstruction performance of the proposed method, we calculated the peak signal-to-noise ratio (PSNR) and structural similarity index (SSIM), defined as:

$$\text{PSNR} = 10 \log_{10}\left(\frac{\max(K)}{\|K - K'\|_2^2}\right), \quad (5)$$

$$\text{SSIM} = \frac{(2\mu_K \mu_{K'} + c_1)(2\sigma_{KK'} + c_2)}{(\mu_K^2 + \mu_{K'}^2 + c_1)(\sigma_K^2 + \sigma_{K'}^2 + c_2)}, \quad (6)$$

where $K$ and $K'$ represent the ground truth (the phantom data) and the target image (the reconstructed image), respectively. max (·) denotes the maximum value. $\mu_K$, $\mu_{K'}$ and $\sigma_K$, $\sigma_{K'}$ are the mean and standard deviations in the square window of images $K$ and $K'$, respectively. $\sigma_{KK'}$ is the covariance of $K$ and $K'$, $c_1 = (0.01L)^2$ and $c_2 = (0.03L)^2$ are constant values, and $L$ is the dynamic range of the ground truth.

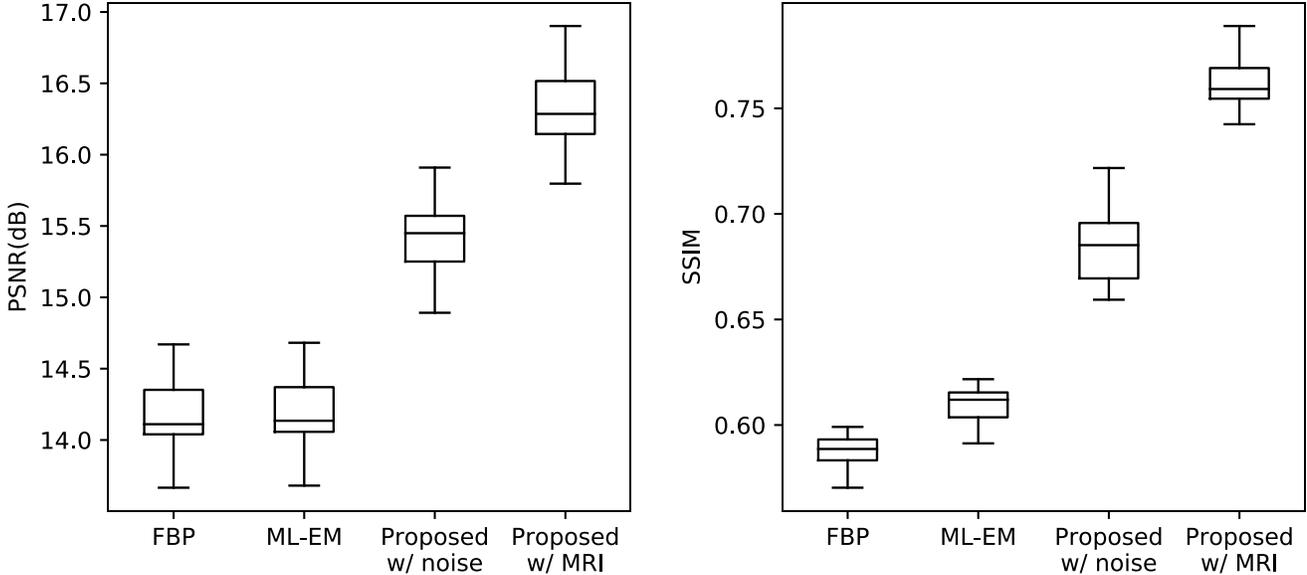

Fig. 3. Quantitative results of the reconstructed images with respect to PSNR (left) and SSIM (right) with different reconstruction algorithms. The line within the box represents the median value, and the upper and lower lines of the box represent the 75-th and 25-th percentiles, respectively. The upper and lower whiskers represent the maximum and minimum values, respectively.

## IV. RESULTS AND DISCUSSION

Fig. 2 shows three orthogonal slices of reconstructed images using different reconstruction algorithms. The proposed direct reconstruction method using random noise and MRI input accurately reconstructed PET images without concealing or losing the more subtle information when compared with both the FBP and ML-EM algorithms. In particular, the proposed method with MRI input recovered more details of the cortical structures, unlike the conventional FBP and ML-EM algorithms. Fig. 3 shows box plots of the PSNR and SSIM for the different reconstruction algorithms. The average quantitative values for the FBP, MLEM, and the proposed method with the random noise and MRI input were 14.16, 14.18, 15.40 and 16.31 dB in PSNRs, and 0.588, 0.611, 0.685 and 0.761 in SSIMs, respectively. The PSNR and SSIM of the proposed direct reconstruction were much higher than those of the FBP and MLEM algorithms. In addition, further improvement of the image quality was achieved using MRI information as the prior input vector. These results demonstrate that the proposed direct reconstruction method yields better performance than other analytical and iterative reconstruction algorithms.

Compared with other deep learning-based PET reconstruction algorithms, our proposed direct reconstruction has a unique approach. A typical deep learning-based direct image reconstruction requires a large number of images that have been reconstructed using conventional image reconstruction methods for network training; however, our proposed method only requires measured projection data. Therefore, the proposed method can help eliminate biases in the datasets, such as artifacts that occur in the image reconstruction process. In addition, our method can be seamlessly applied to different domains, such as different PET scanners, PET tracers, and the detection of previously unknown diseases, because no prior dataset is required. Gong et al. recently proposed to combine iterative PET image reconstruction with the DIP framework to alternately solve the likelihood function in the projection space and the CNN in the image space, using the alternating direction method of multipliers (ADMM) algorithm [28]. Our proposed direct reconstruction was formulated using the parameterization of only the CNN weights for the reconstructed image. Therefore, the proposed method is a novel direct reconstruction framework that uses only the optimization process of the CNN in a single step, unlike the ADMM algorithm which requires three steps.

The main limitation of this study is that we evaluated only the [$^{18}$F]FDG simulation dataset of brain data. In the future, further evaluations with real datasets, other organs, and other PET tracers are needed. Furthermore, the performance of our method in low-dose PET scans needs to be investigated.

## V. CONCLUSIONS

In this study, we proposed a direct PET image reconstruction method incorporating the DIP framework. Our proposed method incorporates a forward projection model with a loss function to achieve unsupervised direct PET image reconstruction from sinograms. Compared with the conventional FBP and ML-EM algorithms, our proposed direct reconstruction method showed superior performance, with improvements of 2 dB, and 0.15 in the PSNR and SSIM, respectively, on the [$^{18}$F]FDG simulation dataset. Future work will further evaluate this method with real datasets and data from other organs.

ACKNOWLEDGMENT

The authors gratefully acknowledge the scientific advice provided by Dr. Hiroyuki Ohba and Dr. Norihiro Harada from the Central Research Laboratory, Hamamatsu Photonics K.K.